\documentclass[11pt,a4paper,english,showpacs,superscriptaddress]{revtex4}

\usepackage{amssymb}
\usepackage{amsmath}
\usepackage{graphicx}
\usepackage{babel}
\usepackage{hyperref}

\begin{document}

\title{Perturbative finiteness of the three-dimensional Susy QED to all orders}

\author{A. F. Ferrari}
\author{M. Gomes}
\author{A. C. Lehum}
\affiliation{Instituto de F\'\i sica, Universidade de S\~ao Paulo\\
Caixa Postal 66318, 05315-970, S\~ao Paulo, SP, Brazil}
\email{alysson,mgomes,lehum,ajsilva@fma.if.usp.br}
\author{A. Yu. Petrov}
\affiliation{Departamento de F\'{\i}sica, Universidade Federal da Para\'{\i}ba\\
 Caixa Postal 5008, 58051-970, Jo\~ao Pessoa, Para\'{\i}ba, Brazil}
\email{petrov@fisica.ufpb.br}
\author{A. J. da Silva}
\affiliation{Instituto de F\'\i sica, Universidade de S\~ao Paulo\\
Caixa Postal 66318, 05315-970, S\~ao Paulo, SP, Brazil}
\email{alysson,mgomes,lehum,ajsilva@fma.if.usp.br}


\begin{abstract}

Within the superfield formalism, we study the ultraviolet properties of the three-dimensional supersymmetric quantum electrodynamics. The theory is shown to be finite at all loops orders in a particular gauge.

\end{abstract}

\maketitle

The presence of divergences is one of the main properties of quantum field theory. This has motivated the development of renormalization methods and the search for special finite field theories. The expectations of finding finite theories are strongly related with supersymmetry, which is well known to improve the ultraviolet behavior of models due to mutual cancellation of bosonic and fermionic contributions. Some notable examples in four dimensions are in the realm of extended supersymmetric theories, such as the ${\cal N}=4$ super-Yang-Mills~\cite{N4S} and some ${\cal N}=2$ superconformal models~(see e.g. \cite{n2sc}), which turn out to be finite (discussions on the existence of noncommutative finite field theories can be found in~\cite{JJ,Khoze:2000sy}). Three-dimensional models have better ultraviolet properties, and are therefore natural candidates to be finite. Indeed, the pure (i.e. without matter) Yang-Mills theory in three dimensions was shown to be finite in~\cite{Sor}. As for supersymmetric models, the pure Yang-Mills-Chern-Simons model was shown to be super-renormalizable and, furthermore, finite, in~\cite{RRN}. Minimally coupled to matter, three-dimensional supersymmetric gauge theories are still super-renormalizable, with superficial divergences appearing up to two-loops. In~\cite{ours}, the noncommutative Abelian and non-Abelian models where shown to be one-loop finite. The remaining problem is the study of the two-loop quantum corrections in these theories, which would allows us to establish if they are finite. As a first step in this direction, in this work we will show the two-loop finiteness of the (commutative) three-dimensional supersymmetric quantum electrodynamics (SQED$_3$), coupled to matter, by explicitly calculating the relevant Green functions.

The starting point of our study is the classical action of SQED$_3$,
\begin{eqnarray}\label{eq1}
S=\int{d^5z}\Big{\{}\frac{1}{2}W^{\alpha}W_{\alpha}
-\frac{1}{2}\overline{\nabla^{\alpha}\Phi}\nabla_{\alpha}\Phi+M\bar\Phi\Phi\Big{\}}~,
\end{eqnarray}

\noindent
where $W^{\alpha}=\frac{1}{2}D^{\beta}D^{\alpha}A_{\beta}$, and $\nabla^{\alpha}=(D^{\alpha}-ieA^{\alpha})$ 
is the gauge supercovariant derivative. Here and further we use the notations and conventions adopted in~\cite{SGRS}. Using the definition of $\nabla^{\alpha}$, we can explicitly rewrite Eq.~(\ref{eq1}) as
\begin{eqnarray}\label{eq1a}
S&=&\int{d^5z}\Big{\{}\frac{1}{2}W^{\alpha}W_{\alpha}
+\bar\Phi(D^2+M)\Phi +
i\frac{e}{2}\big[~D^{\alpha}\bar{\Phi}A_{\alpha}\Phi-\bar\Phi A^{\alpha}D_{\alpha}\Phi~\big]
-\frac{e^2}{2}\bar\Phi\Phi A^{\alpha}A_{\alpha}\Big{\}}~.
\end{eqnarray}

\noindent
This action is invariant under the following infinitesimal gauge transformations:
\begin{eqnarray}\label{eq2}
\delta\bar\Phi=-ie\bar\Phi\,K~,\quad\,
\delta\Phi=ieK\Phi~,\quad\,
\delta A_{\alpha}=-D_{\alpha}K~,
\end{eqnarray}

\noindent
where the gauge parameter $K=K(x,\theta)$ is a real scalar superfield.

The quantization of this theory requires the inclusion in Eq.~(\ref{eq1a}) of the gauge fixing term and the corresponding Faddeev-Popov ghosts action, 
\begin{eqnarray}\label{eq3}
S_{GF+FP}&=&\int{d^5z}\Big[-\frac{1}{4\alpha}
D^{\alpha}A_{\alpha}D^2D^{\beta}A_{\beta}-\bar{c}D^2c\Big]~.
\end{eqnarray}

\noindent
The propagators of the model can be cast as,
\begin{eqnarray}\label{eq4}
\langle \Phi(k,\theta_1)\bar\Phi(-k,\theta_2)\rangle&=&-i\frac{(D^2-M)}{k^2+M^2}\delta_{12}~,\nonumber\\
\langle A_{\alpha}(k,\theta_1)A_{\beta}(-k,\theta_2)\rangle&=&\frac{i}{2}\frac{D^2}{(k^2)^2}
\left(D_{\beta}D_{\alpha}-\alpha D_{\alpha}D_{\beta}\right)\delta_{12}~,\nonumber\\
\langle c(k,\theta_1)\bar{c}(-k,\theta_2)\rangle&=&i\frac{D^2}{k^2}\delta_{12}~,
\end{eqnarray}

\noindent
where $\delta_{12}=\delta^2(\theta_1-\theta_2)$. Note that, as this theory is Abelian and commutative, the ghosts decouple.

To describe the renormalization properties of the model, we must calculate the superficial degree of divergence $\omega$ of an arbitrary diagram. We denote the number of vertices of the form 
$(D^{\alpha}\bar\Phi A_{\alpha}\Phi-\bar\Phi A_{\alpha}D^{\alpha}\Phi)$ and $\bar\Phi\Phi A^2$ by $V_3$ and $V_3$, respectively. The number of propagators for the $\Phi$ and gauge superfields are given by $P_{\Phi}$ and  $P_{A}$. For an arbitrary diagram, the superficial degree of divergence $\omega$ is given by
\begin{eqnarray}\label{eq24}
\omega=2L-2P_A-P_{\Phi}+\frac{V_3}{2}~.
\end{eqnarray} 

\noindent
Indeed, each loop contributes to $\omega$ with $3$ from the integral in $d^3k$ and $-1$ from the contraction of the loop to a point. Each gauge propagator contributes $-2$, and each matter propagator $-1$. The number of propagators in a given Feynman diagram can be written in terms of the number of external superfields ($E$) and vertices as
\begin{eqnarray}\label{eq22}
P_{\Phi}=\frac{1}{2}(2 V_3 + 2 V_3 -E_{\Phi})~,\hspace{1cm}
P_{A}=\frac{1}{2}(V_3+2V_3-E_{A})~.\nonumber
\end{eqnarray} 

\noindent
Using the well known topological identity $L+V-P=1$, we obtain,
\begin{eqnarray}\label{eq25}
\omega=2-V_4-\frac{V_3}{2}-\frac{E_{\Phi}}{2}-\frac{N_D}{2}~,
\end{eqnarray}

\noindent
where $N_D$ is the number of operators $D^{\alpha}$ acting on the external legs of the diagram.

\begin{figure}[htbp]
 \begin{center}
\includegraphics[]{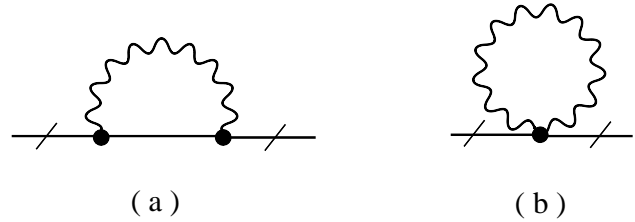}
  \end{center}
\caption{One loop contribution to the two-point function of the scalar superfield. Continuous lines represents the external fields $\Phi$ and $\bar\Phi$, and wave lines represents the gauge superfield propagator.}\label{Fig1}
\end{figure}

\begin{figure}[htbp]
  \begin{center}
\includegraphics[]{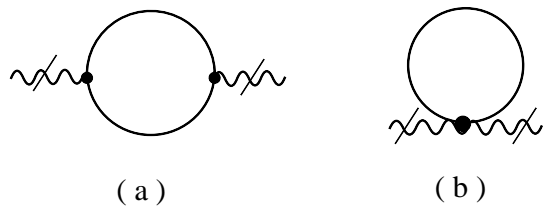}
  \end{center}
\caption{One loop contribution to the two-point function of gauge superfield $A_{\alpha}$. Continuous lines represent the matter superfield propagator, and wave crossed lines represent the external gauge superfield.}\label{Fig2}
\end{figure}

\begin{figure}[htbp]
 \begin{center}
\includegraphics[height=6cm ,angle=0 ,width=12cm]{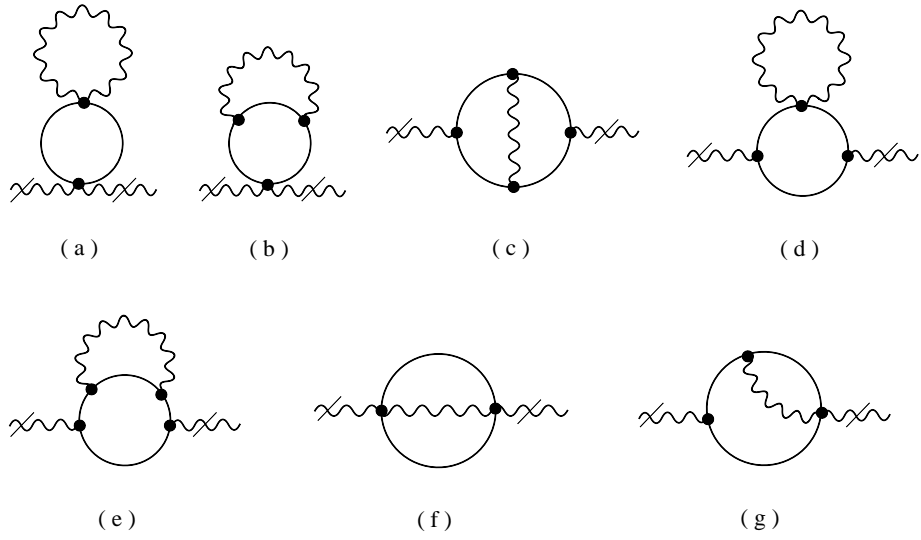}
  \end{center}
\caption{Logarithmically divergent two-point diagrams.}\label{Fig3}
\end{figure}

It follows from Eq.~(\ref{eq24}) that there are no superficially divergent supergraphs at three or higher loop orders, or with more than two external legs. Two point vertex functions can be divergent at one and two loop orders. As for the one loop graphs, the only potentially linearly divergent are those in Fig.~\ref{Fig2}, whose sum we will show to be finite; the two-point vertex function of the $\Phi$ superfield in Fig.~(\ref{Fig1}) happens to be finite by power counting. The logarithmically divergent graphs appear at two loops (Fig.~\ref{Fig3}), and their finiteness will also be established by direct computation. 

The one-loop diagrams that contribute to the two-point vertex function of the $\Phi$ superfield are depicted in the Fig.~\ref{Fig1}. The expression corresponding to the diagram \ref{Fig1}(a) is given by
\begin{eqnarray}\label{eq6}
S_{\Phi\bar\Phi a}&=&\frac{e^2}{8}\int\!\frac{d^3p}{(2\pi)^3}\int\!\frac{d^3k}{(2\pi)^3}
\frac{1}{\left(k^2\right)^2 [(k-p)^2+M^2]}\nonumber\\
&\times&\bar\Phi(p,\theta) \left[
7(1-\alpha)(k\cdot p)(D^2+M)+3\alpha k^2(D^2+M)-7k^2D^2+\alpha Mk^2
\right]\Phi(-p,\theta)~,
\end{eqnarray}

\noindent
whereas the diagram~\ref{Fig1}(b) vanishes since $\int d\theta_1 \, \delta_{12} (D^2)^2 \delta_{12} = 0$. Similarly to~\cite{ours}, the two-point vertex function of the scalar superfield, given by Eq.~(\ref{eq6}), is finite in any gauge, but it takes the simplest form in the Feynman gauge ($\alpha=1$),
\begin{eqnarray}\label{eq7}
S_{\Phi\bar\Phi}&=&\frac{e^2}{2}\int\!\frac{d^3p}{(2\pi)^3}\bar\Phi(p,\theta)(D^2+M)\Phi(-p,\theta)
\int\!\frac{d^3k}{(2\pi)^3}\frac{1}{k^2[(k-p)^2+M^2]}~.
\end{eqnarray}

Two diagrams contributing to the radiative correction to the two-point vertex function of the gauge superfield $A^{\alpha}$ are depicted in Fig.~\ref{Fig2}. The contribution of the diagram~\ref{Fig2}(a) reads
\begin{eqnarray}\label{eq8}
S_{AAa}&=&-\frac{e^2}{2}\int\!\frac{d^3p}{(2\pi)^3}d^2\theta~A^{\alpha}(p,\theta)A_{\alpha}(-p,\theta)
\int\!\frac{d^3k}{(2\pi)^3}\frac{1}{(k^2+M^2)[(k-p)^2+M^2]}\nonumber\\
&\times& \Big{\{}(k^2+M^2)A^{\alpha}(p,\theta)A_{\alpha}(-p,\theta)
+\frac{M}{2}A^{\beta}(p,\theta)(p_{\beta}^{\alpha}-\delta_{\beta}^{\alpha}~D^2)A_{\alpha}\\
&-&\frac{1}{4}A^{\beta}(\delta^{\alpha}_{\beta}~p^2+p_{\beta}^{\alpha}~D^2)A_{\alpha}(-p,\theta)\Big{\}}~.\nonumber
\end{eqnarray}

\noindent
while, for the diagram \ref{Fig2}(b), we have
\begin{eqnarray}\label{eq8a}
S_{AAb}=\frac{e^2}{2}\int\!\frac{d^3p}{(2\pi)^3}d^2\theta~A^{\alpha}(p,\theta)A_{\alpha}(-p,\theta)
\int\!\frac{d^3k}{(2\pi)^3}\frac{1}{k^2+M^2}~.
\end{eqnarray}

\noindent
Adding up Eqs.~(\ref{eq8},\ref{eq8a}), and carrying out some algebraic manipulations, we arrive at 
\begin{eqnarray}\label{eq9}
S_{AA}&=& e^2 \int\!\frac{d^3p}{(2\pi)^3}d^2\theta~f(p,M)\Big{\{}W^{\alpha}W_{\alpha}-MA^{\alpha}W_{\alpha}\Big{\}}~,
\end{eqnarray}

\noindent
where
\begin{eqnarray}\label{eq10}
f(p,M)=\int\frac{d^3k}{(2\pi)^3}\frac{1}{(k^2+M^2)[(k-p)^2+M^2]}~.
\end{eqnarray}

It is important that the linear divergent terms present in Eqs.~(\ref{eq8}) and (\ref{eq8a}), that would correspond to the generation of a mass for the gauge superfield, cancel among themselves, so the gauge invariance of the result is manifest. The correction $S_{AA}$ turns out to be finite, similarly to~\cite{ours,cpn}, and corresponds to non-local contributions to the Maxwell and Chern-Simons terms. 

Now let us turn to the two-loop approximation. Up to now, we have shown that the logarithmic divergences in the model can arise only at the two-loop level. One should notice that the logarithmic divergences are absent in the one-loop order in all three-dimensional field theories due to the symmetry of the Feynman integral~\cite{ours} (however, this is not so in theories with effective dynamics obtained within the ${1}/{N}$ expansion~\cite{cpn,sig}). Since we are interested in the divergent parts, and keeping in mind Eq.~(\ref{eq25}), we will explicitly calculate all contributions to the effective action proportional to $A^{\alpha}A_{\alpha}$, where no covariant derivatives end up in the external legs, in the two-loop approximation. The corresponding supergraphs are depicted in Fig.~\ref{Fig3}. 

The calculational procedure adopted by us was the following: the D-algebra manipulations on the two-loops supergraphs were performed with the help of the Mathematica$^{\copyright}$ package SusyMath~\cite{Ferrari:2007sc}. From the resulting (unintegrated) terms, we extracted all those  proportional to $A^{\alpha}A_{\alpha}$. Finally, we considered the lowest term in an expansion of this result around vanishing external momentum, which corresponds to a mass term for the gauge superfield $A^\alpha$,
\begin{eqnarray}\label{eq27}
S_{AA(mass)}&=&\int\frac{d^3p}{(2\pi)^3}d^2\theta~\Gamma_{AA}~A^{\alpha}(p,\theta)A_{\alpha}(-p,\theta)~.
\end{eqnarray}

\noindent
Any ultraviolet divergence present at the two-loop level must appear in this term.

We may now describe the outcomes of our calculations. The diagram \ref{Fig3}(a) and \ref{Fig3}(d) happens to vanish identically, as a consequence of the vanishing of \ref{Fig1}(b). As for the remaining diagrams, we obtained
\begin{equation}
\Gamma_{AA} = \Gamma_{AA(b)}+\Gamma_{AA(c)}+\Gamma_{AA(e)}+\Gamma_{AA(f)}+\Gamma_{AA(g)}\,,
\end{equation}

\noindent
where,
\begin{eqnarray}\label{eq26b}
\Gamma_{AA(b)}&=&i\frac{e^4}{6}\int\frac{d^3k}{(2\pi)^3}\frac{d^3q}{(2\pi)^3}
\Big[-\frac{(1-\alpha)~(k\cdot q)}{(k^2+M^2)[(k+q)^2+M^2](q^2)^2}\nonumber\\
&-&\frac{1}{(k^2+M^2)[(k+q)^2+M^2]q^2}
+\frac{2M^2(1+\alpha)}{(k^2+M^2)^2[(k+q)^2+M^2]q^2}\Big]~,
\end{eqnarray}
\begin{eqnarray}\label{eq26c}
\Gamma_{AA(c)}&=&i(1-\alpha)\frac{e^4}{64}\int\frac{d^3k}{(2\pi)^3}\frac{d^3q}{(2\pi)^3}
\Big[\frac{4~(k\cdot q)^2~k^2}{(k^2+M^2)^2[(k+q)^2+M^2]^2 (q^2)^2}\nonumber\\
&-&\frac{2~(k^2)^2}{(k^2+M^2)^2[(k+q)^2+M^2]^2q^2}
+\frac{3(k\cdot q)^2M^2}{(k^2+M^2)^2[(k+q)^2+M^2]^2 (q^2)^2}\nonumber\\
&-&\frac{M^2k^2}{(k^2+M^2)^2[(k+q)^2+M^2]^2q^2}
\Big]~,
\end{eqnarray}
\begin{eqnarray}\label{eq26e}
\Gamma_{AA(e)}&=&i\frac{e^4}{16}\int\frac{d^3k}{(2\pi)^3}\frac{d^3q}{(2\pi)^3}
\Big[\frac{(1-\alpha)~(k\cdot q)^2}{(k^2+M^2)^2[(k+q)^2+M^2](q^2)^2}\nonumber\\
&-&\frac{8~M^4~\alpha}{(k^2+M^2)^3[(k+q)^2+M^2]q^2}
+\frac{k^2}{(k^2+M^2)^2[(k+q)^2+M^2]q^2}\nonumber\\
&+&\frac{7\alpha~(k^2)^2}{(k^2+M^2)^3[(k+q)^2+M^2]q^2}
-\frac{\alpha~k^2~M^2}{(k^2+M^2)^3[(k+q)^2+M^2]q^2}
\Big]~,
\end{eqnarray}
\begin{eqnarray}\label{eq26f}
\Gamma_{AA(f)}&=&-i(1+\alpha)\frac{e^4}{2}\int\frac{d^3k}{(2\pi)^3}\frac{d^3q}{(2\pi)^3}
\frac{1}{(k^2+M^2)[(k+q)^2+M^2]q^2}~,
\end{eqnarray}
\begin{eqnarray}\label{eq26g}
\Gamma_{AA(g)}&=&-i\frac{e^4}{6}\int\frac{d^3k}{(2\pi)^3}\frac{d^3q}{(2\pi)^3}
\Big[\frac{2(1-\alpha)(k\cdot q)}{(k^2+M^2)[(k+q)^2+M^2](q^2)^2}\nonumber\\
&+&\frac{(1-2\alpha)(k\cdot q)^2}{(k^2+M^2)^2[(k+q)^2+M^2](q^2)^2}
+\frac{k^2+2(1-\alpha)M^2}{(k^2+M^2)^2[(k+q)^2+M^2]q^2}
\Big]~.
\end{eqnarray}

\noindent
The two-loop integrals were performed in the dimensional reduction scheme, using formulas from~\cite{Dias:2003pw}, and we obtained
\begin{eqnarray}\label{eq28}
\Gamma_{AA}=\frac{ie^4}{384\pi^2}\Big\{ (\alpha+8)\left[\frac{1}{\epsilon}- (\gamma-\ln (4\pi)-1)-\ln\left(\frac{2M}{\mu}\right)\right]+\frac{3}{8}(7\alpha-3)\Big\}~,
\end{eqnarray}

\noindent
where $\gamma$ is the Euler's constant. Differently from what happens in one-loop, the mass term for the gauge superfield does not vanish identically. This fact signalizes that our regularization is not preserving the gauge symmetry at two loops~\cite{RRN,Siegel:1980qs}. Gauge symmetry may be restored by the introduction of a mass counterterm in the classical Lagrangian. For the specific gauge $\alpha=-8$, the two-point vertex function turns out to be finite, and only a finite counterterm is needed to ensure the Ward identities.

We have studied the perturbative finiteness of the three-dimensional supersymmetric quantum electrodynamics. The only possible divergence in the theory, arising in the two-point vertex function of the gauge superfield, turns out to vanish for a specific gauge choice $\alpha=-8$. This fact was established by means of a direct calculation of the potentially divergent vertex functions, up to the two loop order. The finiteness of the $n$-point functions only in a specific gauge also happens in other supersymmetric models, such as in the ${\cal N}=4$ super-Yang-Mills theory in four spacetime dimensions~\cite{N4S}. It is interesting to contrast our results with the ones in~\cite{RRN} where, in the absence of matter, the Yang-Mills-Chern-Simons model was shown to be finite. Note that, in the component formalism used by~\cite{RRN}, ultraviolet divergences can appear up to three loop order, whereas in the superfield formalism used by us, they appear at the most at the two loop order. A peculiarity of the dimensional reduction regularization scheme is that a finite mass counterterm is needed to ensure the gauge invariance of the vertex functions. It is natural to expect that the non-Abelian generalization of this theory will also display two-loop finiteness, up to some possible restrictions, similarly to the one-loop commutative and noncommutative situations~\cite{Sor,ours}. As a final remark, we would like to point out that a natural extension of our work would be the evaluation of the two loops quantum corrections in three dimensional noncommutative gauge theories, which we have already studied at the one loop level in~\cite{ours}. There, the momentum-dependent trigonometric factors arising from the Moyal product would be an additional complication.

\vspace{1cm}

{\bf Acknowledgements.} A.C.L. would like to thank W. Siegel for useful comments. This work was partially supported by Funda\c{c}\~{a}o de Amparo \`{a} Pesquisa do Estado de S\~{a}o Paulo (FAPESP) and Conselho Nacional de Desenvolvimento Cient\'{\i}fico e Tecnol\'{o}gico (CNPq). The work by A. F. F. has been supported by FAPESP, project 04/13314-4. The work by A. Yu. P. has been supported by CNPq-FAPESQ DCR program, CNPq project No. 350400/2005-9.


\end{document}